# Spontaneous creation and annihilation dynamics and strain-limited stability of magnetic skyrmions

Frederic Rendell-Bhatti[1*], Raymond J. Lamb[1], Johannes W. van der Jagt[2], Gary W. Paterson[1], Henk J. M. Swagten[2] and Damien McGrouther[1]

[1]SUPA, School of Physics and Astronomy, University of Glasgow, Glasgow G12 8QQ, UK
[2]Department of Applied Physics, Eindhoven University of Technology, 5612 AZ Eindhoven, The Netherlands

*f.rendell.1@research.gla.ac.uk

Magnetic skyrmions are topological magnetic spin structures exhibiting particle-like behaviour. They are of strong interest from a fundamental viewpoint and for application, where they have potential to act as information carriers in future low-power computing technologies. Importantly, skyrmions have high physical stability because of topological protection. However, they have potential to deform according to their local energy environment. Here we demonstrate that, in regions of high exchange energy density, skyrmions may exhibit such extreme deformation that spontaneous merging with nearest neighbours or spawning new skyrmions is favoured to attain a lower energy state. Using transmission electron microscopy and a high-speed imaging detector, we observe dynamics involving distinct configurational states, in which transitions are accompanied by spontaneous creation or annihilation of skyrmions. These observations raise important questions regarding the limits of skyrmion stability and topological charge conservation, while also suggesting a means of control of skyrmion creation and annihilation.

**Introduction**

Skyrmions were originally proposed as a soliton model of the nucleon by Tony Skyrme in 1962[1], but they have since been observed as emergent topological quasiparticles in a variety of condensed matter systems including superconductors[2], liquid crystals[3] and magnetic thin films[4]. Although the specific microscopic mechanisms leading to the formation of skyrmion structures differ in each case, the emergent topological structures bear many similarities and skyrmions are therefore of fundamental interest. Topology defines the distinctness of the geometries in continuous systems such as vector fields, as the inability to continuously map one such system to another. Skyrmions can be understood as localised topological knots within continuous fields that cannot be unwound without a discontinuous break in their structure. This property is fundamental to the idea of topological protection, equipping skyrmions with a degree of stability against perturbations and contributing to the description of skyrmions as quasiparticles. Skyrmions can be categorised according to the topological charge defined by their structure, whereby processes in which they participate are expected to obey topological charge conservation. Understanding the real-world physical stability and possible interactions of these topological structures is of great importance, especially in magnetic thin films, in which magnetic skyrmions are expected to play a role in future low power computing technologies[5].

In non-centrosymmetric helimagnetic materials, such as B20 FeGe, the lack of inversion symmetry in the crystal structure, combined with strong spin-orbit coupling gives rise to competition between the Dzyaloshinskii–Moriya interaction (DMI) and the Heisenberg exchange interaction[6]. Owing to this competition, materials with a significant DMI will host a helical magnetic ground state, defined by continuously rotating magnetisation along a particular direction with an associated helical wavevector[7]. It is possible to describe the emergence of a hexagonal magnetic skyrmion lattice crystal (SkX) as a magnetic phase transition involving the nucleation of discrete topological quasiparticles from the helical ground state under favourable conditions of magnetic field and temperature[8]. The description of skyrmions as particle-like objects is supported by experimental observations that indicate their high stability[9] and their ability to organise themselves into domains within a SkX[10] (as commonly seen in atomic systems) or even microcrystals analogous to colloidal crystallisation[11]. Additionally, the inherent rigidity of individual skyrmions allows current-induced propagation via spin-transfer torque without destruction[12]. It has been reported that the transition of a SkX into the topologically trivial ferromagnetic state (high field) or helical state (low field) involves magnetic singularities known as Bloch points that fulfil topological charge conservation during these transitions[13,14]. However, owing to the extreme difficulties associated with performing nanometre resolved magnetic imaging of a stochastic dynamic process taking place in the range of durations from nanoseconds to tens of microseconds, it remains to be experimentally determined whether topological charge is conserved during skyrmion annihilation processes.

Here, we report the direct observation of spontaneous, repeatable creation and annihilation of individual skyrmions in FeGe using Lorentz transmission electron microscopy and a high-speed imaging detector with a temporal resolution of 10 ms (see Methods for details). Creation and annihilation processes occurring on timescales of tens of milliseconds are observed as a repeated lateral skyrmion motion, captured (Supplementary Movie 1) across a SkX domain boundary separating lattice domains. These thermally driven transitions involve a number of configurational states with distinct lifetimes, separated by relative energy barriers and characterised by skyrmions of five- and seven-fold coordination (5-7 defects). Aided by micromagnetic simulations, we show that these transitions occur in regions of localised high energy density, dominated by exchange energy, while the wider SkX remains in the stable ground state under conditions of constant applied field and temperature. Furthermore, the observed transitions involve discrete changes in topological charge, and we propose that extreme deformation of key skyrmions leads to the emergence and subsequent destruction of antiskyrmions, fulfilling topological charge conservation. This demonstrates an example of magnetic strain-limited stability for skyrmions and suggests that topological protection may be overcome through the spontaneous emergence of topological structures that provide lower energy pathways for skyrmion creation or annihilation.

## Results

**Identifying and characterising configurational states**

Transitions between six unique configurational states (Fig. 1b–g) were observed within the region defined by the dashed box in Fig. 1a. These states are further categorised into three primary states, namely, $P_1$, $P_2$ and $P_3$ (Fig. 1b–d), and three transition states, namely, $T_1$, $T_2$ and $T_3$ (Fig. 1e–g). Each of the identified states features a lattice defect structure composed of skyrmions of five- and seven-fold coordination (5–7 defects), highlighted in Fig. 1b by the conjoined pentagon and heptagon, structurally analogous to the 5–7 defects found in graphene[15]. Note, only the overall outline of the 5–7 defect is shown in later images. The skyrmions at the centre of the pentagons and heptagons, herein denoted as SkCoP and SkCoH, respectively, differ structurally from regular lattice skyrmions which possess hexagonal symmetry[16] by being more spatially compressed (under five-fold coordination) or expanded (under seven-fold coordination)[17]. The motion of the 5–7 defects changes the position of the domain boundary (shown by the blue/red/green dotted lines in Fig. 1a) and is accompanied by very slight shifts in the regular lattice skyrmions in the immediate vicinity as they accommodate the change.

In order to analyse the order of occurrence, each frame was assigned to one of the observed $P_x$ or $T_x$ states (see Methods for details). A portion of this sequence data can be seen in Fig. 1h. The system was observed to transition reversibly between $P_x$ states via $T_x$ states, an example of the reversibility of the transitions is recorded in section J of Fig. 1h. However, some $P_x$–$P_x$ transitions observed did not visibly appear to proceed via a $T_x$ state (see section K in Fig. 1h). In these cases, we assume that, owing to its short-lived nature compared to the frame acquisition time, occupation of the $T_x$ state simply comprised a small fraction of the duration of an individual frame. Evidence for such an assumption comes from observations in which some smearing of intensity suggested multiple states ($P_x$ and $T_x$) being captured in a single frame and will be discussed further below. Furthermore, the system did not always transition between different $P_x$ states, as recorded in section L of Fig. 1h. This pattern suggests that, once the system is in a $T_x$ state it can lower its energy by transitioning to a neighbouring $P_x$ state or back to its starting $P_x$ state. Additionally, the transition probabilities (Supplementary Fig. 1) demonstrate that the system does not randomly transition between the six states but rather hops between adjacent $P_x$ states via the appropriate $T_x$ state. Finally, we describe these transitions in terms of skyrmion creation and annihilation. During $P_x$ to $T_x$ transitions, a SkCoP is created, appearing to be formed by the division of an expanded skyrmion of seven-fold coordination, SkCoH. Conversely, in a $T_x$ to $P_x$ transition, a skyrmion of five-fold coordination, SkCoP, is annihilated by merging with a nearest-neighbour skyrmion. Thus, skyrmion creation is associated with $P_x$ to $T_x$ transitions, and skyrmion annihilation is associated with $T_x$ to $P_x$ transitions. In $T_x$ states there is higher local skyrmion density than in $P_x$ states, which, through there being a reduction of the mean inter-skyrmion distance, we infer there to be an increase in local energy density. The precise structure of the skyrmions within the 5–7 defects is elucidated through the spatial analysis of the six observed states. A series of four successive frames of the $P_1$ state (Fig. 2a) shows that the SkCoH appears to have distinct curvature around its centre and that this curvature increases as the sequence progresses. This behaviour demonstrates the dynamic nature of the SkCoH structure, in which the skyrmion appears to be deforming between elongated and dumbbell structures. Such structures represent a significant deformation from the regular hexagonally coordinated lattice skyrmions. Each pair of images in Fig. 2b,c, consists of a summed intensity image across all observed examples of the state (left) and a Delaunay triangulation analysis of the average skyrmion positions based on their geometric centre of mass (right). In Fig. 2d,e, integrated intensity profiles (extracted from the dotted box in each summed intensity image in Fig. 2b,c) are plotted. They record the intensity values along a path joining two skyrmions of regular six-fold coordination, passing through a SkCoH in the case of $P_x$ states (Fig. 2d) or through a SkCoP in the case of $T_x$ states (Fig. 2e).

A key feature of the $P_x$ states is the relatively deformed structure of the SkCoH in the 5–7 defect (Fig. 2b). From the summed intensity images, the deformed structure appears to possess the aforementioned dumbbell shape on average. This observation is supported by the intensity profiles in Fig. 2d, which show additionally that the SkCoH in each of the $P_x$ states has a highly similar form. The intensity profiles are characterised by two clearly defined central intensity peaks of approximately equal height, with an intensity 60%–70% that that of regular neighbouring skyrmions of six-fold coordination, and spatial extent double that of regular neighbouring skyrmions of six-fold coordination. Figure 2d also shows that the same-row nearest-neighbour distances (i.e., from the intensity peaks of the SkCoH dumbbell to the intensity peaks of the adjacent regular hexagonally coordinated skyrmions) approximately match those of the bulk lattice: 80 nm. By comparing the SkCoH structure of the two most common states ($P_1$ and $P_2$), it was observed that 41% (205 of 499 frames showing $P_1$) and 87% (328 of 376 frames showing $P_2$) were characterised by a SkCoH skyrmion possessing an intensity central double peak, as described for the summed intensity images in Fig. 2b. This observation suggests that the precise structure of the SkCoH fluctuates, but a common state (indeed the most common state for the less stable $P_2$ state) is the dumbbell-like form.

Following a similar set of analyses for the $T_x$ states, the SkCoP in the 5–7 defect appears to be less well defined in the summed intensity images appearing somewhat smeared (Fig. 2c). The intensity profiles in Fig. 2e across the three skyrmions involved show three defined intensity peaks, with a central peak intensity approximately 75%–90% that of the neighbouring peaks. Between their positions, the intensity decreases to around 58%–68% of the central peak values, highlighting the transient nature of the $T_x$ states. The compressed nature of the SkCoP in the $T_x$ state can also be observed in both the Delaunay triangulation map of Fig. 2c and in the intensity profile of Fig. 2e, in which same-row nearest-neighbour distance is measured to be 64 nm, 20% closer than in the bulk lattice. This deviation from the equilibrium inter-skyrmion distance corresponds to an increase in energy density (see Supplementary Fig. 2b).

A non-time varying 5–7 defect (from a different location along the same SkX domain boundary) is shown in Fig. 2f, with the intensity across the SkCoH shown in the integrated line profile. Again, the SkCoH is more spatially extended than regular hexagonally coordinated lattice skyrmions, but, instead of the two distinct central intensity peaks that were observed for the dynamic SkCoHs in Fig. 2b, there is a single peak with a shoulder. This pattern indicates that this stationary SkCoH more likely possesses a single core, displaced toward one side. The distance between the two intensity peaks belonging to regular skyrmions in Fig. 2f is approximately 192 nm, whereas this same distance is around 208 nm in the $P_x$ states in Fig. 2d. This indicates that the SkCoHs involved in 5–7 defect dynamics have a greater lateral spatial extent, giving rise to the additional deformation observed that yields the dumbbell structure. This behaviour may be a contributing factor for the skyrmion creation process through modification of the energy landscape. Comparing the dynamic and stationary SkCoH integrated intensity line profiles (Fig. 2d and Fig. 2f, respectively), the lower relative central peak intensity and greater spatial extent suggests higher variance in the dynamic SkCoH structure. Fresnel (defocused) Lorentz transmission electron microscope (TEM) imaging, while providing high-contrast imaging of the skyrmions, does not possess sufficient resolution to allow the identification of the detailed magnetic structure within the deformed skyrmions (ref. 8). However, it is possible to correlate the Fresnel images with higher-resolution differential phase contrast images of 5–7 defects, as will be discussed presently.

By comparing the Delaunay triangulation images in Fig. 2b,c, it is clear that the triangular cells joining each skyrmion location have a lower area in the $T_x$ states, making the overall nearest-neighbour distances shorter than in the $P_x$ states (shown by the colour scale). This difference reinforces the fact that skyrmion creation causes regions of higher skyrmion density ($T_x$ states) that reduce local energetic stability, leading to later relaxation to a $P_x$ state through annihilation of a skyrmion. The fact that these transitions occur spontaneously suggests that the relative energies of the six states are close and that the thermal energy fluctuations of the system are sufficient to overcome the potential energy barriers separating them.

**Relative stability and energy landscape**

In order to investigate the energetics of the system, the lifetimes of each state over the 1000 frames were measured. Figure 3a shows the lifetime distributions along with the total number of observations of each of the six states, indicating that the $P_1/T_1$ and $P_2/T_2$ states were the most commonly observed. Owing to the short-lived nature of the $T_x$ and $P_3$ states, these transitions may not always be captured individually, but may instead be averaged across the frame duration as previously mentioned. This explanation also suggests why $T_x$ states were not always observed during transitions, since (for example) state lifetimes of 1 ms would contribute only 10% intensity to the final 10 ms frame. It has been established that the system can transition to or from the same $P_x$ state ($P_1$–$T_1$–$P_1$ for example). However, if the $T_x$ state was very short-lived, then it is possible that the system could appear to have stayed in the $P_x$ state ($P_1$–$P_1$). This summing effect might explain some observations of very long lifetimes (around 400 ms) for the $P_1$ state in Fig. 3a.

The timescale of a magnetic transition between two states, facilitated by the thermal energy of the system, while separated by a simple potential energy barrier, can be deduced from the Néel–Arrhenius equation:

$$\frac{1}{\tau} = \nu_0(T) e^{-\frac{\Delta E}{kT}} \qquad (1)$$

where the lifetime $\tau$ of a particular state is the product of the attempt frequency $\nu_0(T)$ (number of transition attempts per second) and an exponential term containing the Boltzmann factor, which gives the probability of a transition with energy barrier $\Delta E$ at temperature $T$. By assuming that the observed transitions obey equation (1) to a first approximation and using an attempt frequency of $10^9$ Hz and $kT$ of 0.022 eV (see Methods for details), relative energy barriers for the skyrmion creation and annihilation processes were calculated (Fig. 3b). The most energetically stable configuration is the $P_1$ state (with a barrier height on the order of 0.55 eV), corresponding to its relatively long lifetime. The less stable $P_2$ state has 28% more observations than $P_1$, this disparity can be understood by the fact that the $P_2$ state is both energetically and spatially positioned between the $P_1$ and $P_3$ states. This suggests that there may be entropy-compensation effects involved, due to the increased number of possible transition state configurations accessible from the $P_2$ state (decreasing its observed lifetime)[18,19].

By assuming that each transition is not affected by past transitions, the transition behaviour can be modelled using a Markov chain. The observed transition probabilities and average lifetimes of each state were combined to give the schematic in Fig. 3c. In this diagram only transitions which occurred more than once are included and the arrows(circles) are scaled to the relative transition probabilities(lifetimes). During a transition from $P_1$ to $P_3$, it is more probable for the system to proceed in multiple lower energy steps via the $P_2$ state instead of a single step. This multi-step process is demonstrated by the individual transition probabilities given in Fig. 3c where zero transitions were observed between the $P_1$ and $T_3/P_3$ states directly (see Supplementary Fig. 1 for the full transition matrix). Instead, every state has a connection both to and from the $P_2$ state (red circle). It should be noted that the $T_1$ and $T_2$ states are labelled at the same position in the energy landscape (Fig. 3b). This is because although they are spatially distinct (Fig. 2c), they differ only by a small shift in the skyrmion positions local to the SkCoP and therefore were approximately equally likely to occur. However, Fig. 3c demonstrates that there is a small probability bias when transitioning from a $T_x$ state to the corresponding $P_x$ state. For example, since $T_2$ to $P_2$ requires a smaller overall shift in skyrmion positions it is more probable than $T_2$ to $P_1$.

**Skyrmion creation and annihilation mechanisms**

In order to gain insight into the energy terms that might govern the skyrmion creation and annihilation processes we performed a series of micromagnetic simulations of both isolated 5–7 defects and those within a SkX boundary (see Supplementary Note 1). We found that the stretched core of the SkCoH is a region of high energy density due to an extended region of magnetisation antiparallel to the applied perpendicular magnetic field (providing a large positive Zeeman energy contribution). The magnitude of this increase in energy was found to be determined by the degree of deformation (spatial extent) of the SkCoH. Regions of increased exchange energy density around the SkCoP were observed due to the reduction from the equilibrium of the inter-skyrmion distance. Thus, it follows that the splitting of an expanded SkCoH ($P_x$ to $T_x$) lowers the Zeeman energy but in creating an extra skyrmion, leads to regions of high skyrmion density ($T_x$ states) with steeper magnetisation gradients, increasing the local exchange energy. Therefore, these states quickly collapse back into one of the $P_x$ states via skyrmion annihilation ($T_x$ to $P_x$), lowering the skyrmion density but resulting in significantly deformed SkCoHs with a high Zeeman energy contribution. We propose that the interplay and balance of these two energy configurations results in the observed repeated, spontaneous transitions involving skyrmion creation and annihilation. Due to limitations in the temporal resolution of the LTEM measurements, the fine details of the transition mechanisms are not yet accessible experimentally. However, informed by micromagnetic simulations (see Methods and Supplementary Note 2 for details), we propose mechanistic pathways involving antiskyrmions for both creation and annihilation of skyrmions through splitting and merging processes.

In figure 4 we take account of the topological charges associated with skyrmions and suggest mechanisms that explain how our experimentally observed processes may be accomplished. Fig. 4a shows how a significantly deformed SkCoH with a topological charge of $N = -1$ involved in the observed dynamic processes may split into two separate skyrmions ($P_x$ to $T_x$ transition). This first occurs through stretching of its form from an elongated structure to a dumbbell structure (as shown experimentally in Fig. 2a). Such a deformation is continuous, and consequently, there is no associated change in topology or topological charge. However, at the centre of the dumbbell-like SkCoH structure, significant magnetisation gradients and curvature occur, creating a region of locally high energy density and low energetic stability. This leads to formation of two skyrmion structures with a total topological charge of $N = -2$, connected by a central region containing an antiskyrmion with $N = +1$ (positive topological charge since it has the same core polarity as the two formed skyrmions). Thus, the antiskyrmion fulfils topological charge conservation during the SkCoH splitting into two skyrmions. Since the middle state in Fig. 4a could not be captured experimentally, we investigated its creation using micromagnetic simulations, through enforced splitting of an elongated skyrmion using magnetic field gradients (Supplementary Fig. 3e). The resulting transition state can be seen in Fig. 4c, where two individual skyrmions are joined by an antiskyrmion. The final step involves the unstable antiskyrmion reducing in size, until eventually the central spin rotates in-plane and the antiskyrmion topology is destroyed, accompanied by a step-change in total topological charge of $N = -1$ to $N = -2$ (Supplementary Fig. 3b right panel). This antiskyrmion collapse mechanism has been proposed theoretically by Desplat et al (ref 19), where the saddle point configuration corresponds to the state preceding the central spin flipping (Fig. 4c right panel).

Fig. 4b demonstrates the case for skyrmion merging ($T_x$ to $P_x$ transition). We propose that the reverse mechanism is followed such that a reduction in proximity of two merging skyrmions with a total topological charge of $N = -2$ leads to a singled elongated skyrmion with $N = -1$, with a trapped region of magnetisation parallel to the external applied field. This is an antiskyrmion of opposite polarity to the previous splitting case (Fig. 4a) and thus has an associated topological charge of $N = -1$. Therefore, the antiskyrmion again fulfils the role of topological charge conservation during skyrmion merging. Support for such a mechanism again came from micromagnetic simulations where applied magnetic field gradients caused merging of two neighbouring skyrmions (see Supplementary Fig. 3d). The resulting transition state and antiskyrmion corresponding to the middle state in Fig. 4b can be seen in Fig. 4d. Again, the antiskyrmion reduces in size until it is eventually destroyed when the central spin flips, resulting in a step change in topological charge from $N = -2$ to $N = -1$ (Supplementary Fig. 3b left panel). The reverse mechanism is similar to one suggested for skyrmion lattice inversion during polarity switching of an applied external magnetic field, whereby antiskyrmions are formed at the point of skyrmion core merging[20].

This notion of antiskyrmions conserving topological charge during creation or annihilation events has also been studied in dipolar magnets, in which antiskyrmions have a higher stability than that in chiral magnets[21]. The thickness of our sample (on the order of the FeGe helical length, 70 nm) implies that variation in magnetisation through the thickness will be minimal. This leads us to believe that the mechanism involves the creation and annihilation of antiskyrmions rather than the involvement of Bloch point zippers, as discussed in ref. 13, which studied a bulk sample. However, both of these mechanisms involve topological structures that are created and eventually destroyed to fulfil topological charge conservation during transitions. The mechanisms involving antiskyrmions demonstrated in Fig. 4 also bear resemblance to the proposed mechanisms of magnetic vortex core reversal, whereby antivorticies are created and subsequently destroyed[22].

Recently, a theoretical study using a saddle point search method was used to determine saddle point configurations of skyrmions undergoing duplication and collapse[23]. These configurations bear many similarities with the dumbbell-like structures observed and are suggestive of the proposed transition state configurations involving antiskyrmions. It has been shown that, using a 360° domain wall model, skyrmion deformation does not significantly lower the universal skyrmion energy (i.e., energy required to reduce skyrmion radius to zero)[24]. This value should be sufficient to prevent spontaneous skyrmion annihilation processes at the temperatures studied here. However, we propose that thermal fluctuations combined with extreme skyrmion deformation lead to the formation of the antiskyrmions as described. Thereby providing new energy pathways for skyrmion creation and annihilation that are much lower than the calculated universal skyrmion energy.

To better understand the magnetisation distribution within deformed SkCoH/SkCoP, we performed comparisons with other 5–7 defects not varying in time (i.e., stationary under observation). Figure 4e shows a static 5–7 defect from a different SkX imaged using the higher spatial resolution technique of differential phase contrast from four-dimensional scanning transmission electron microscopy (4D STEM) (see Methods for details). The images show that the SkCoH is quite significantly deformed compared to the lattice skyrmions surrounding it. For the magnitude of the in-plane component of magnetic induction (Fig. 4e bottom panel), the SkCoH exhibits a significant deformation with clearly defined strip-like regions of in-plane magnetisation connecting it to two neighbouring skyrmions, isolating out of plane regions as suggested in Fig. 4b. Additionally, it is possible to identify a stretched core region and some kinking of the magnetisation circulating around it. This situation bears similarity to the structure suggested by the Fresnel images of the SkCoH seen for the dynamic 5–7 defects (Fig. 2) but does not possess the deformation required for splitting at the temperatures studied.

## Discussion

We observed spontaneous dynamic processes involving the creation and annihilation of individual skyrmions at SkX domain boundaries. Frame-based analysis allowed us to identify key states in the process and to observe variations in the structure of the SkCoH /SkCoP in 5–7 defects that leads to skyrmion creation or annihilation. Naturally, the processes observed are governed by thermal fluctuations and induced exploration of a relative potential energy landscape and appeared at first sight not to fulfil the concept of topological charge conservation. However, aided by micromagnetic simulations and based on our high-spatial-resolution imaging of other deformed but time-non-varying SkCoH skyrmions, we have proposed that skyrmions can exhibit extreme deformation, leading to the formation of antiskyrmions. Through subsequent random local magnetisation fluctuations, the antiskyrmion object may be destroyed, and subsequently lead to lower energy pathways involving neighbouring skyrmions merging or separating.

Our observations provide evidence of magnetic strain-limited skyrmion energetic stability in lattices, i.e., based on Zeeman and exchange interaction energies. This finding is contrary to many reports of high energetic stability, arising from topological protection for both isolated skyrmions and those in lattices[24,25,26,27]. Our observations provide the potential for a new method of controlling skyrmion creation and annihilation through engineering SkX boundaries with high magnetic strain, induced via the intentional patterning of magnetic/non-magnetic defects with key dimensions comparable to SkX periodicity. Although this study was limited to investigating a single SkX boundary, we have generally observed that 5-7 defect density and the rate of skyrmion dynamic events increase as a function of SkX misorientation angle[28]. Further time-resolved imaging studies of the extremely deformed skyrmions involved in these dynamic processes may provide new insights into the dynamic transformation processes of topological structures within condensed matter systems.

## Methods

### Lorentz Fresnel TEM Imaging

Sections of single-crystal FeGe were extracted from a bulk host crystal using the in-situ liftout technique in a focused ion beam-scanning electron microscope (FIB-SEM) instrument (FEI Nova 200 Nanolab). The extracted sections possessed a (110) normal and thickness <100 nm in order to provide sufficient electron transparency. In the TEM (JEOL ARM200cF), the sample was cooled using liquid nitrogen in a Gatan HC3500 specimen holder. For Fresnel mode, defocused imaging of the skyrmion state was performed at a temperature of 253 K with the TEM operating in Low Mag (objective lens OFF mode). Out-of-plane magnetic fields were applied by partially exciting the objective lens. States containing multiple Skyrmion crystal lattices, and therefore boundaries, were induced by starting from a disordered helical state and then by applying and continually increasing the strength of the applied magnetic field up to a value of 510 Oe. Direct filming of the dynamics occurring at skyrmion lattice boundaries was performed using a high-speed, direct illumination, single electron counting 256 × 256 pixel imaging detector (Quantum Detectors Merlin for EM 1R). In these experiments, filming was performed using an exposure time of 10 ms and in continuous read/write mode (i.e., no time gaps between frames), yielding a frame rate of 100 fps.

### Lorentz DPC Imaging

DPC imaging of skyrmions, stationary under observation, was performed also on a JEOL ARM200cF operating in a custom LM-STEM mode[29]. An electron probe with semi-convergence angle 415 mrad and full-width-half-maximum of 6 nm was scanned across the sample. 4D STEM acquisition was performed over a 256 x 256 pixel scan array, 1 ms pixel dwell time (using the same pixelated single electron counting detector as for the Fresnel imaging) to acquire a dataset that was subsequently processed using DPC phase-correlation algorithm in the fpd library[30]. Isotropic smoothing was applied to the resultant image in order to reduce noise effects and to make clear the features of the distorted skyrmion structure.

### Image Analysis

Assignment of each image frame to one of the configurational states was performed by first using the centre of mass of intensity of the skyrmions involved in the lateral motion to create Delaunay triangulation maps. Then, visual inspection and integrated intensity line-profile measurements allowed the identification of the six unique observed states. A Delaunay triangulation of each individual frame was then categorised into one of the six observed states. All analysis of Fresnel images shown in Fig. 2 first involved a background subtraction using the rolling ball algorithm. The images shown in Fig. 2 are the result of the summation of many frames in order to significantly reduce influence from the level of statistical noise associated with the limited electron counts in the short movie frames.

### Assumptions for relative energy barrier calculations

A precise value of the temperature-dependent attempt frequency $v_0(T)$ in Eq. 1 can be calculated only from detailed information about the minimum energy path taken during transitions[31], the exact mechanism of which has not yet been determined for skyrmion merging and splitting. However, it has been reported that the attempt frequency for magnetic skyrmion annihilation in thin films is around $10^9 - 10^{10}$ Hz[32], which is in the range typically reported for other magnetic systems[33]. Theoretical studies into the minimum energy pathway of isolated skyrmion collapse also report strong entropy compensation effects under certain conditions, particularly metastable skyrmions far from equilibrium conditions leading to variations in $v_0(T)$ of up to five orders of magnitude.[19,34] However, we justify the use of an attempt frequency of $10^9$ Hz since we are only seeking to calculate relative barrier heights. Additionally, in the present study we do not have experimental indication of the appropriate modification to $v_0(T)$. Finally, any error in the estimated value of $v_0(T)$ will simply shift the entire energy landscape vertically, leaving the relative barrier heights unchanged.

**Micromagnetic simulations**

Micromagnetic simulations were performed using a finite difference (FD) approach, utilising the well-established MuMax3 code[35]. Material parameters were chosen in order to represent the estimated material parameters of FeGe[36], they were $D$ = 1.58 mJ m$^{-2}$, $A$ = 8.78 pJ m$^{-1}$ and $M_s$ = 384 kA m$^{-1}$. Further details on simulation parameters can be found in Supplementary Notes 1 and 2. The purposes of the simulations were firstly to estimate the main energy contributions involved in 5-7 defect structures and secondly to gain insight into the possible mechanistic pathways of skyrmion splitting and merging. Therefore, we chose to perform all simulations at zero temperature as we were not seeking to reproduce the dynamics observed experimentally. In order to enforce splitting and merging of individual skyrmions a strong localised magnetic field gradient was imposed. This field gradient acted as a proxy for the magnetic strain across a boundary, forcing skyrmions to merge or split as observed experimentally. Therefore, the results may only give approximations to the true pathway. We chose ultra-thin films at zero temperature in order for the simulations to be valid, given the thickness-temperature-field phase space of B20 FeGe[20]. Finally, the micromagnetic simulations were performed to support insight into the mechanistic pathways which are currently inaccessible experimentally due to timescales.

## Data availability

The Lorentz Fresnel, DPC and micromagnetics data that support the findings of this study are available in the Enlighten: Research Data repository (DOI: 10.5525/gla.researchdata.1030) (ref. 37).

**Acknowledgements**

We would like to thank Y. Kousaka and Y. Togawa of Osaka Prefecture University, Japan for provision of the FeGe single crystals and the EPSRC-JSPS Core to Core grant (EP/M024423/1) and EPSRC/SFI CDT in Photonic Integration and Advanced Data Storage (EP/S023321/1) for financially supporting the work.

**Author Contributions**

F.R.B. performed Fresnel TEM data analysis, carried out micromagnetic simulations and prepared the manuscript. R.J.L. and J.W.J were responsible for preparation of FeGe samples and for acquisition of Fresnel TEM and DPC STEM data sets and analysis. G.W.P. performed advanced processing of the 4D STEM DPC data and assisted with drafting the manuscript. H.J.M.S and D.M. jointly supervised the work. D.M. was responsible for conception and design of the work and assisted with drafting the manuscript. All authors commented on the manuscript.

**Competing Interests**

The authors declare no competing interests.

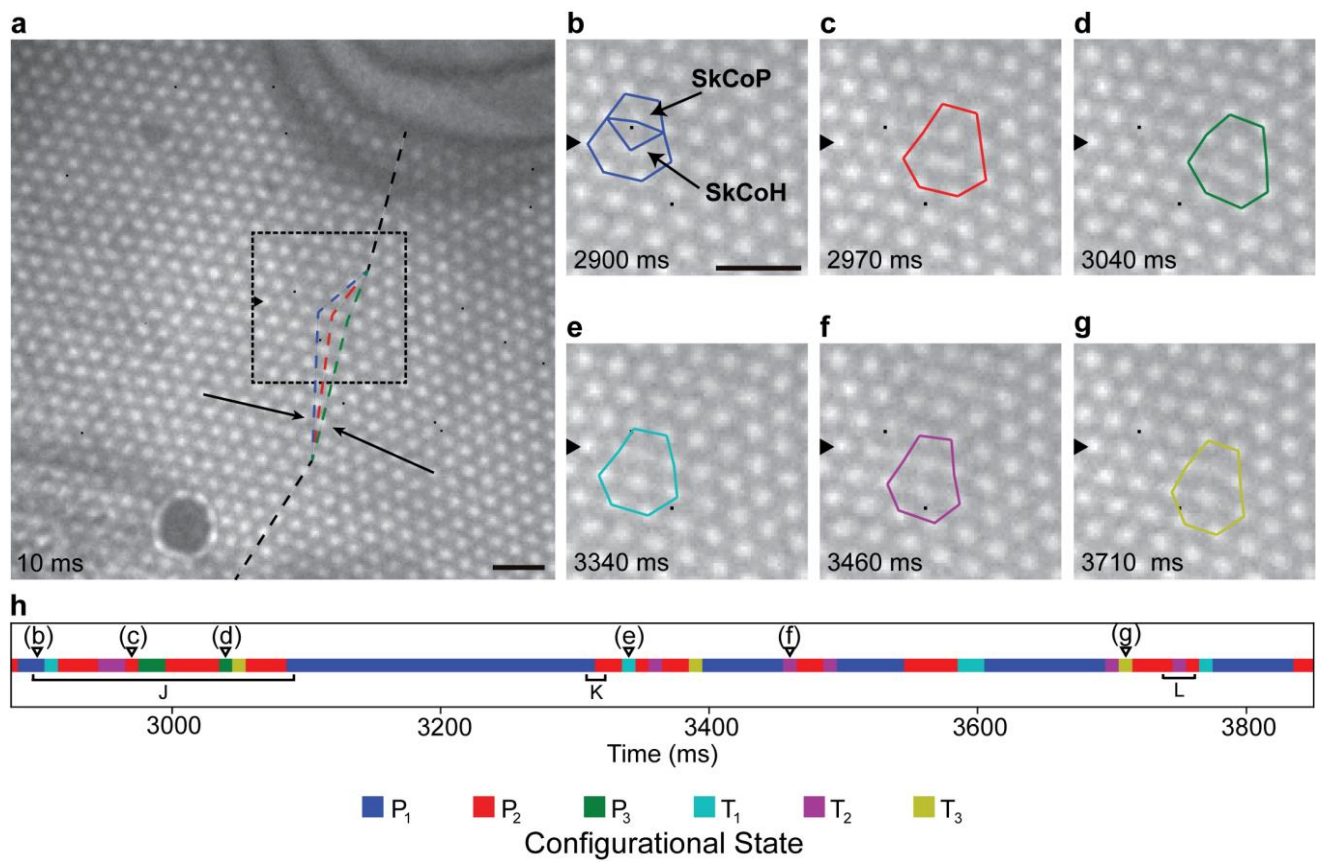

**Fig. 1 | Fresnel TEM images of SkX and identification of configurational states. a**, Entire region of observed skyrmion lattice, showing the domain boundary with an average misorientation angle of 14° (black dashed lined) and principle lattice vectors (arrows). The three coloured (blue, red and green) dashed lines demonstrate how the domain boundary shifts during transitions between the $P_x$ states and the associated lateral skyrmion motion. Scale bar, 200 nm. **b-g**, Examples of the six configurational states observed, with the time of each frame given in the bottom left corner of each image. The black arrowheads point along the row of skyrmions in which lateral motion was taking place (same position in each image). Skyrmions with seven- and five-fold coordination (SkCoH and SkCoP, respectively) are highlighted by the conjoined pentagons and heptagons. Scale bar, 200 nm. **h**, A portion of the transition data showing the progression of reversible transitions between states shown in **b-g**. The markers (b)-(g) additionally indicate the time of occurrence of the images shown in **b-g**. Colour in each panel is associated with the identified configurational states, according to the bottom key.

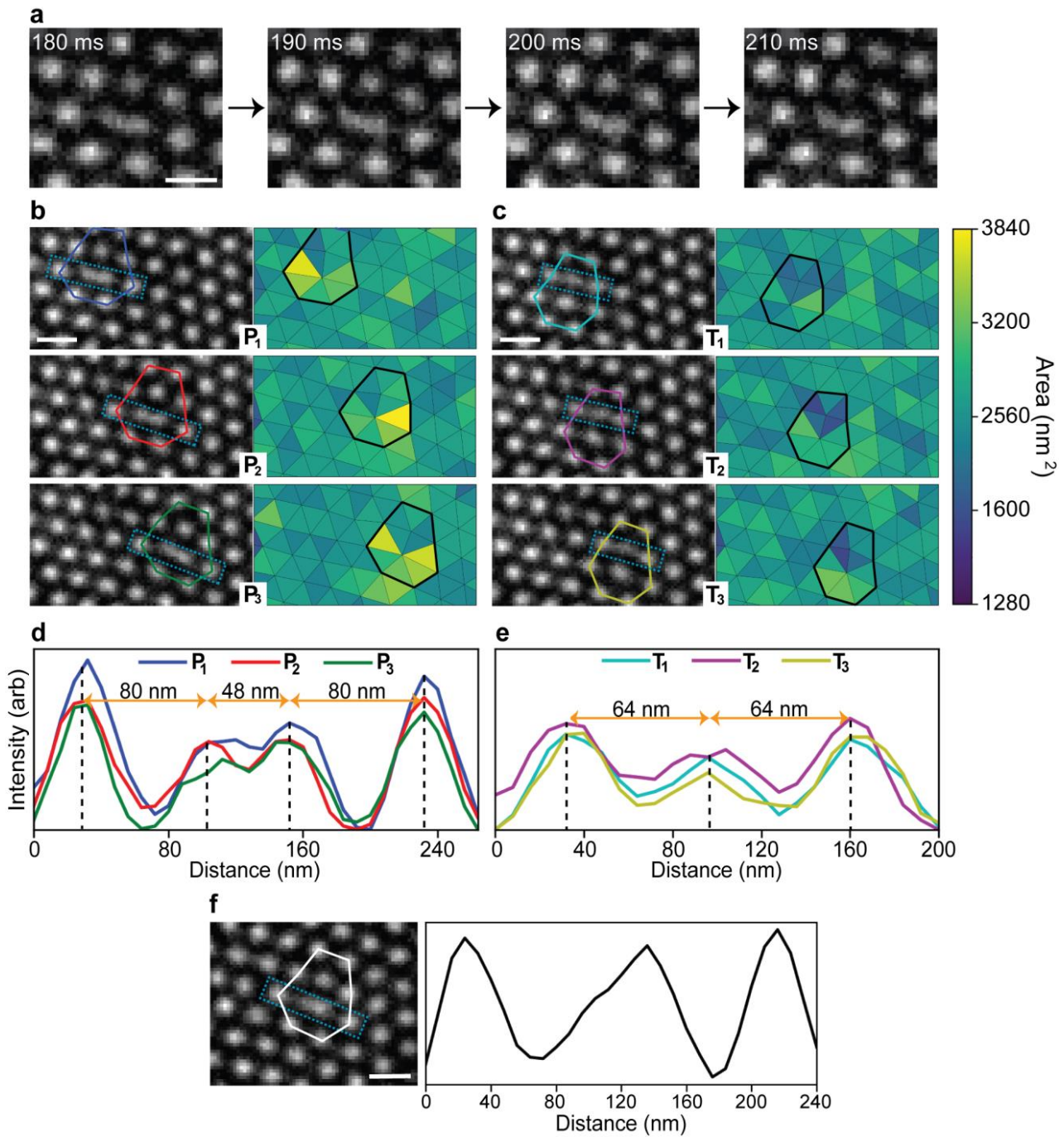

**Fig. 2 | Spatial analysis of the six observed configurational states. a**, Successive frames showing the variation in structure of the SkCoH in the $P_1$ state with the time of each frame given in the bottom right of each image. Scale bar, 100 nm. **b,c**, Summed intensity images of each observed state with coloured outlines (same colour key introduced in Fig. 1) around the 5–7 defect (left) and their Delaunay triangulation calculated from the skyrmion geometric centre of mass (right). The colour bar indicates the area of the triangles in $nm^2$. Scale bars, 100 nm. **d,e**, Integrated line profile plots of the $P_x$ and $T_x$ states respectively, measured across the regions shown in **b,c**, (dashed rectangles). The two outermost peaks are associated with skyrmions of regular six-fold coordination. The double inner peak in **d** demonstrates the dumbbell-like structure of each of the SkCoH in the $P_x$ states. **f**, Summed intensity image and integrated line profile plot across a stationary (under observation) SkCoH, for comparison. This 5-7 defect is located further down the SkX boundary, just below the right black arrow in Fig. 1a. Scale bar, 100 nm.

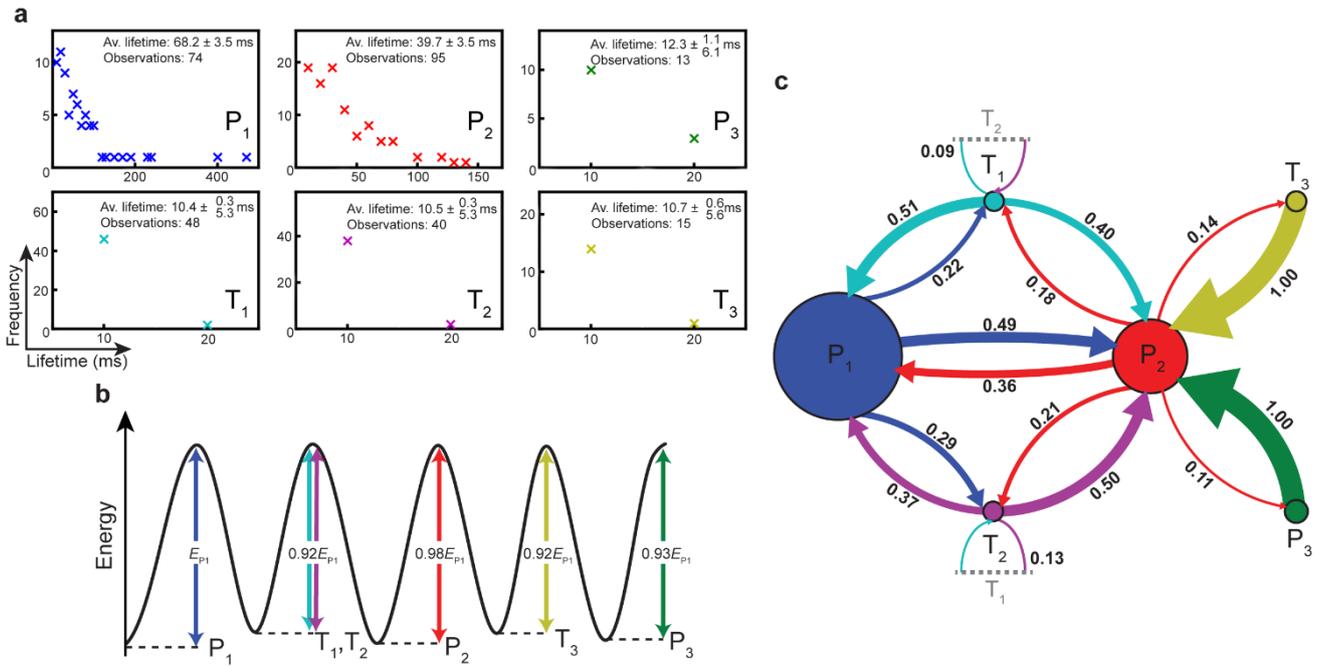

**Fig. 3 | Lifetime distributions, associated energy barriers and transition probabilities of each state. a**, Scatter plots of lifetime distributions for each of the observed configurational states. Half the frame time (5 ms) was added to the lower bound in $T_x$ and $P_3$ lifetime uncertainties due to the short-lived nature of these states not always being captured individually per frame. **b**, Relative energy barriers calculated using the Néel-Arrhenius equation and the average lifetime associated with each state. Each barrier height is given relative to the most stable state ($P_1$), which had an energy barrier height on the order of 0.55 ev. **c**, Observed state transitions represented as a Markov chain process, only including transitions observed more than once. Circles(arrows) representing each state(transition) are scaled to the relative lifetimes(transition probabilities).

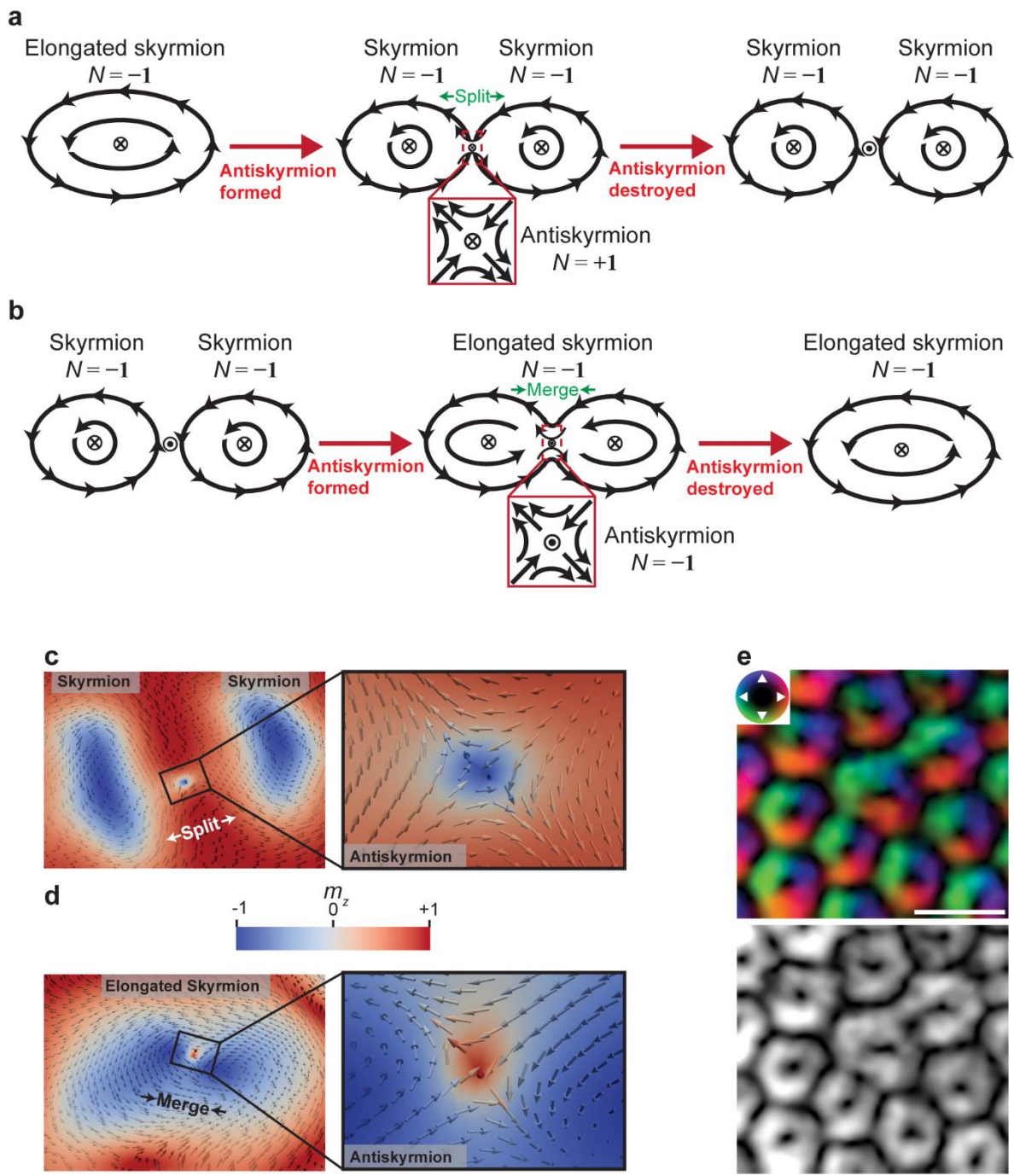

**Fig. 4 | Skyrmion creation and annihilation mechanisms and investigation of the deformed SkCoH structure. a,b**, Proposed mechanisms for skyrmion creation and annihilation via splitting and merging, respectively. $N$ gives the topological charge of each structure involved. **c,d** Micromagnetic simulation snapshots of splitting and merging transition states, closely resembling the middle states in **a,b**, respectively. Colour bar indicates the out of plane ($m_z$) component of magnetisation. **e**, DPC images of a stationary 5–7 defect showing pronounced deformation of the SkCoH structure and indication of merging with nearest neighbour skyrmions. Colour (top) is given by the inset colour wheel which indicates the direction of the in-plane magnetisation, where black indicates out of plane magnetisation. Scale bar, 100 nm.

# Supplementary Information

**Supplementary Figure 1: Transition Matrix**

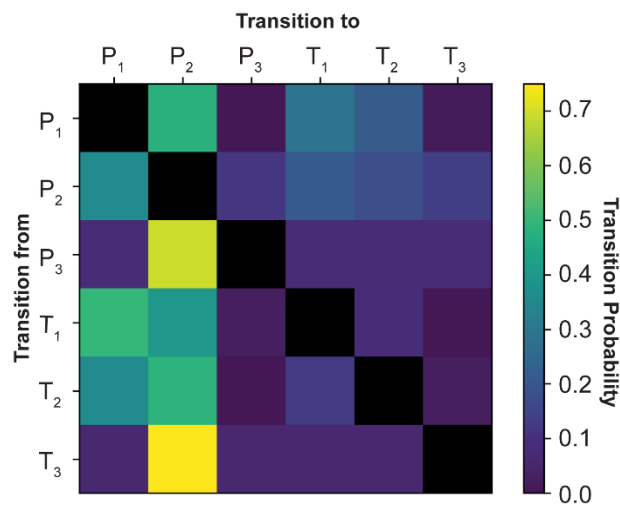

**Supplementary Fig. 1 | Transition matrix of all observed transition probabilities.** Transition matrix with probabilities normalised to unity across rows. This matrix highlights the involvement of the $P_2$ state with most transitions due to its position both geometrically and energetically in the observed transitions. Additionally, the $P_3$ and $T_3$ states transitioned to almost exclusively the $P_2$ state. Colour bar indicates the transition probability of each transition.

**Supplementary Note 1: Micromagnetic simulations of skyrmion lattice boundary**

Investigation of the relative energetic stability of 5–7 defects within a domain boundary was performed using finite difference micromagnetic simulations (MuMax3[1]) involving a pair of isolated 5-7 defects and those within a SkX domain boundary. In order to compare the relative energy densities of a 5-7 defect and regular 6-fold coordinated skyrmions, the appropriate equilibrium lattice parameter was found by varying the dimensions of a SkX triangular unit cell and minimising the energy with an external field of 0.2 T applied out-of-plane [2]. Supplementary Fig. 2a,b shows the relationship between total energy density, $\eta_{tot}$, and skyrmion lattice parameter, $a_{sk}$. The minimum in Supplementary Fig. 2a corresponds to the equilibrium skyrmion lattice parameter, approximately 91 nm, which was the lattice parameter used for the following simulations. The energy density plots in Supplementary Fig. 2c,d (averaged over thickness) were obtained by simulating a thin film of FeGe using periodic boundary conditions in the XY-direction. The simulation parameters for outputs in Supplementary Fig. 2 were $B_z$ = 0.2 T and a damping parameter of $\alpha$ = 0.1 with material parameters $D$ = 1.58 mJ m$^{-2}$, $A$ = 8.78 pJ m$^{-1}$ and $M_s$ = 384 kA m$^{-1}$ in order to represent the material parameters of FeGe[3]. The initial simulation in Supplementary Fig. 2c (left) involves a repeating array of 64 skyrmions, with simulation dimensions 632 nm x 730 nm x 50 nm and a cell size of 1 nm$^3$. Subsequently, a single skyrmion was deleted using a strong localised field and allowed to relax into the state shown on the right. The SkCoHs show up clearly as regions of high energy density due to their elongated forms.

In order to achieve the bi-domain state shown in Supplementary Fig. 2d, the system was initialised in a randomly magnetised state and allowed to relax into a disordered helical state at zero applied magnetic field. A perpendicular magnetic field of 0.4 T was subsequently applied in order to nucleate skyrmions and allowed to evolve for a total simulation time of 1 μs. This length of time was required for the randomly distributed skyrmions to form a skyrmion lattice. This resulted in a multi-domain skyrmion lattice. However, the skyrmions were extremely confined and did not exhibit the hexagonal symmetry observed experimentally,[4] or the equilibrium lattice parameter calculated in Supplementary Fig. 2a. At this point, the perpendicular anisotropy constant was removed, the external field was reduced to 0.2 T. Simulation dimensions of 1670 nm x 980 x 50 nm, corresponded to a minimum energy density and an equilibrium skyrmion lattice parameter of approximately 91 nm. As with the isolated 5-7 defects, the SkCoHs within the boundaries in Fig. S2d can be identified through their elongated cores, corresponding to high total energy density. Additionally, the DMI+exchange energy density plot highlights the increased energy density associated with the compressed SkCoP due to deviation from the equilibrium skyrmion lattice parameter. Exchange and DMI are combined here, because of how MuMax3 calculates the exchange energies.

The relative energies of 5-7 defects when compared to regular 6-fold coordinated skyrmions was calculated for both the simulations shown in Supplementary Fig. 2c,d. The integrated energy was found within the regions enclosed by the dashed white lines and then divided by the total simulation volume enclosed by these regions. The energy density of 5-7 pairs calculated here vary between 0.4% and 0.7% less energetically stable than the 6-fold coordinated skyrmions. The difference is due to the energy associated with the SkCoH core, in Supplementary Fig. 2c they are more deformed (approximately 6% higher *XY*-spatial extent) relative to the regular skyrmions, and thus have a higher maximum energy associated with them (see the $\eta_{tot}$ energy density scales).

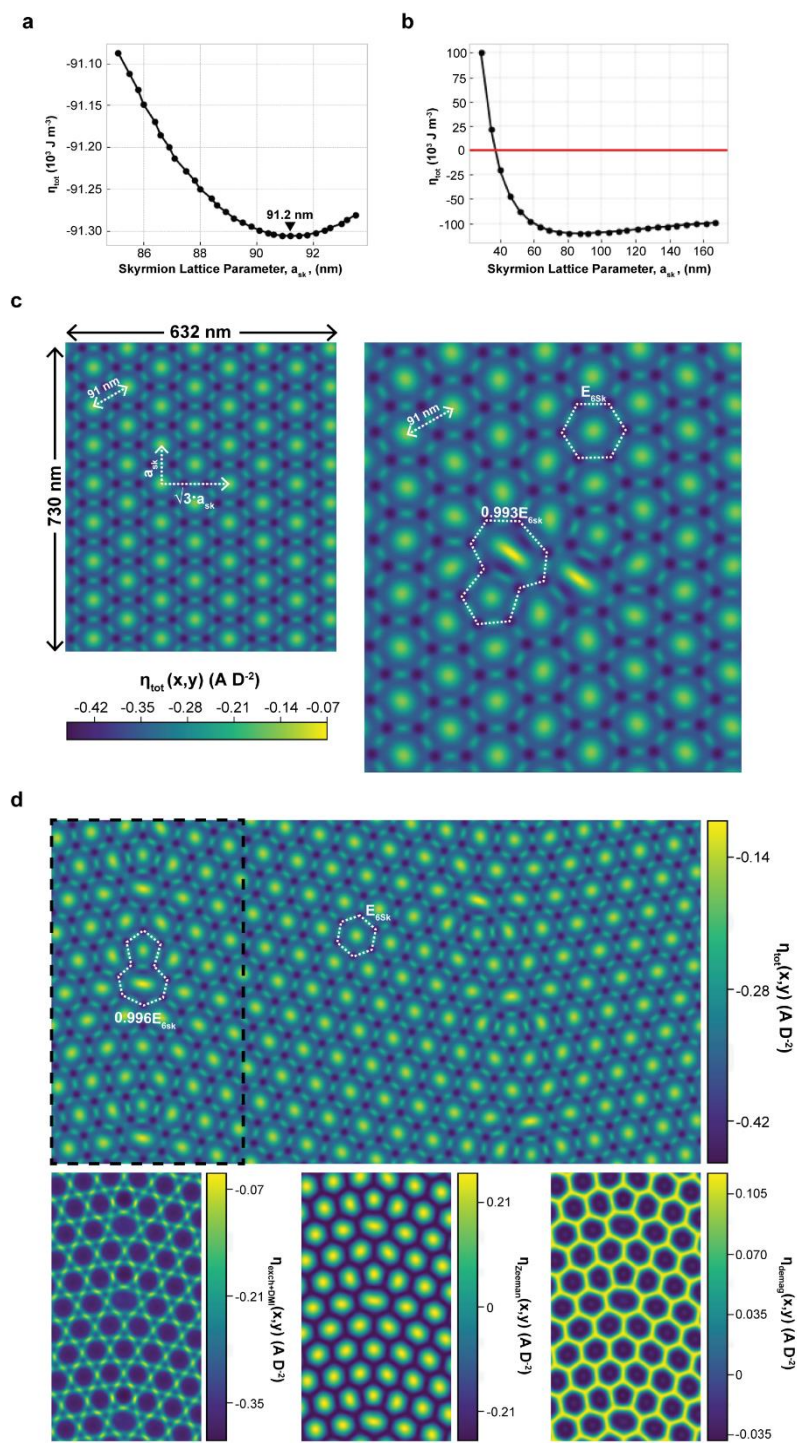

**Supplementary Fig. 2 | Micromagnetic simulations involving 5-7 defects. a,b,** Total energy density plots as a function of skyrmion lattice parameter, $a_{sk}$, for a thickness of 50 nm. Equilibrium lattice parameter of 91 nm was used in future simulations. **c**, Total energy density of continuous SkX (left) and with a pair of 5-7 defects imposed through skyrmion deletion (right). Energy density comparison of a 5-7 defect with regular 6-fold coordinated skyrmion calculated from the regions enclosed by the dashed white lines. **d**, Total energy density of SkX region containing two boundaries with three 5-7 defects in each. Energy density comparison again from the regions enclosed by the dashed white lines. Exchange+DMI, Zeeman and demagnetising energy density plots are from the left-most SkX boundary, enclosed by the black dashed box. A 0.2 T out-of-plane field is applied in all cases and energy densities are given in units of $A/D^2$.

**Supplementary Note 2: Micromagnetic simulations of skyrmion creation/annihilation mechanisms**

By using spatially confined short-duration magnetic field gradients, it was possible to initiate the splitting of an elongated skyrmion (Supplementary Fig. 3a top panel) and the merging of two neighbouring skyrmions (Supplementary Fig. 3a bottom panel). Supplementary Fig. 3d,e shows a series of magnetisation snapshots of skyrmion annihilation and creation, respectively. The colour is given by Supplementary Fig. 3c, where in-plane magnetisation is given by the colour and out-of-plane magnetisation is given by the shade where white points out of the page and black points into the page. The simulation parameters were $D$ = 1.58 mJ m$^{-2}$, $A$ = 8.78 pJ m$^{-1}$ and $M_s$ = 384 kA m$^{-1}$. A damping parameter of of $\alpha$ = 0.02 was used in order to investigate the dynamical aspects of the simulation under study here. A cell volume of 1 nm$^3$ was used, and the sample dimensions were 484 nm × 560nm × 2 nm for the merging case and 512 nm × 256 nm × 2 nm for the splitting case. The merging process was also carried out with a 10 nm thick sample in order to investigate the details of antiskyrmion destruction, and is shown in Supplementary Fig. 3f.

    The merging simulation is shown in Supplementary Fig. 3d, a very short-pulsed (0.7 ns) magnetic field gradient (25 MT m$^{-1}$) was localised around a skyrmion in a hexagonal lattice using a two-dimensional gaussian with a standard deviation of 50 nm in the x-direction and 35 nm in the y-direction (shown by the dashed ellipse in the left panel of Supplementary Fig. 3a). This caused the skyrmion localised under the field gradient to move towards a neighbouring skyrmion and eventually merge. The merging of the two skyrmions results in continuous rotation of magnetisation (black circular arrow) and gives the appearance of an elongated skyrmion with an antiskyrmion superimposed at the centre (the trapped out of plane region). This antiskyrmion subsequently reduces in size, as shown in the bottom set of panels in Supplementary Fig. 3d. Once the antiskyrmion cannot be reduced in size any further, the central spin rotates in plane and the antiskyrmion topology is destroyed (this can be seen clearly in the second panel of Supplementary Fig. 3f). This is accompanied by a large discontinuous change in the topological charge (left panel in Supplementary Fig. 3b) which accounts for the removal of the $N$ = -1 associated with the antiskyrmion. The splitting mechanism is shown in Supplementary Fig. 3e, again a short-pulsed (0.15 ns) magnetic field gradient was localised around the centre of an elongated skyrmion. The standard deviation of the gaussian in this case was 25 nm in both the x- and y-direction. In this case, the elongated skyrmion quickly separates into two skyrmions each with $N$ = -1 and a central antiskyrmion with $N$ = +1 (since it has opposite polarity to the antiskyrmion involved in Supplementary Fig. 3d). This conserves the initial topological charge of $N$ = -1 of the single skyrmion. As before, the antiskyrmion first reduces in size and is then destroyed when the central spin rotates in plane. This is accompanied by a discontinuous change in topological charge (right panel in Supplementary Fig. 3c) corresponding to the removal of the $N$ = +1 associated with the antiskyrmion.

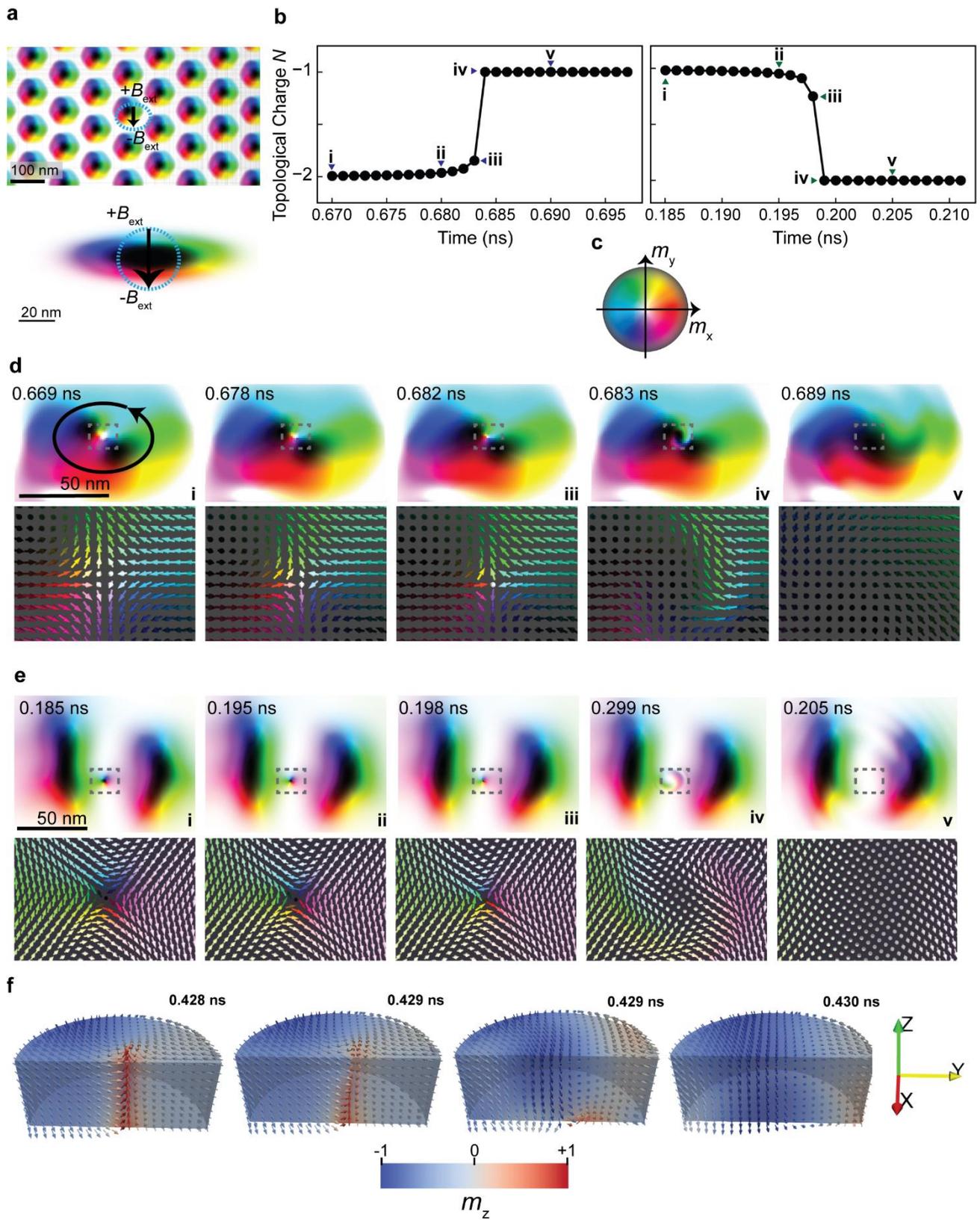

**Supplementary Fig. 3 | Micromagnetic simulations of skyrmion creation/annihilation mechanisms. a**, Initial states of skyrmion annihilation (top) and creation (bottom). Dashed ellipses give the first standard deviation of the field-pulse gaussians, with the arrow showing the direction of the magnetic field gradient from positive to negative field. **b**, Total topological charge as a function of time during the annihilation (left) and creation (right) simulations, labels i-v correspond to the panels in **d** and **e**. **d,e**, A series of scalar plot snapshots of magnetisation (top) and vector plots of the antiskyrmion (bottom) of the skyrmion annihilation/creation mechanisms respectively, colour given by **c**. **f**, Annihilation of an antiskyrmion through a thickness of 10 nm during the merging of two skyrmions.